\documentclass[a4paper,preprint]{revtex4}
\usepackage{epsf}
\usepackage[dvips]{graphicx}
\usepackage{graphics}

\begin{document}
\setcounter{page}{1}
\title{The Kardar-Parisi-Zhang Equation with Temporally Correlated Noise - A Self Consistent Approach}
\author{Eytan Katzav}
\email{eytak@post.tau.ac.il}
\author{Moshe Schwartz}
\email{mosh@tarazan.tau.ac.il} \affiliation {School of Physics and
Astronomy, Raymond and Beverly Sackler Faculty of Exact Sciences,
Tel Aviv University, Tel Aviv 69978, Israel}

\begin{abstract}
In this paper we discuss the well known Kardar Parisi Zhang (KPZ) equation driven by temporally correlated
noise. We use a self consistent approach to derive the scaling exponents of this system. We also draw general
conclusions about the behavior of the dynamic structure factor $\Phi_q(t)$ as a function of time. The approach
we use here generalizes the well known self consistent expansion (SCE) that was used successfully in the case of
the KPZ equation driven by white noise, but unlike SCE, it is not based on a Fokker-Planck form of the KPZ
equation, but rather on its Langevin form. A comparison to two other analytical methods, as well as to the only
numerical study of this problem is made, and a need for an updated extensive numerical study is identified. We
also show that a generalization of this method to any spatio-temporal correlations in the noise is possible, and
two examples of this kind are considered.
\end{abstract}

\maketitle

\section{Introduction}
Nonequilibrium surface growth processes often exhibit a
phenomenon called kinetic roughening, where the surface develops
a self-affine morphology \cite{barabasi95}. Much attention has
been given to a special class of models (ballistic deposition,
Eden, or polynucleation growth), which are described by the
Kardar-Parisi-Zhang (KPZ) equation \cite{kpz86}
\begin{equation}
\frac{{\partial h\left( {\vec r,t} \right)}}{{\partial t}} = \nu
\nabla ^2 h + \frac{\lambda }{2}\left( {\nabla h} \right)^2  +
\eta \left( {\vec r,t} \right)
 \label{1},
\end{equation}
where $h\left( {\vec r,t} \right)$ is the local height of the surface above a $d$-dimensional substrate in a
$\left( {d + 1} \right)$ -dimensional space, $\lambda $ characterizes the tilt dependence of the growth
velocity, $\nu$ is an effective surface tension, and $\eta \left( {\vec r,t} \right)$ is a noise term.

Solutions of eq. (\ref{1}) exhibit scaling behavior. The simplest
quantity to investigate is the surface width $W\left( {L,t}
\right)$ that scales as (see ref. \cite{Family85})
\begin{equation}
W\left( {L,t} \right) = \frac{1}{{\sqrt L }}\left\langle {\sum\limits_{\vec r} {\left[ {h\left( {\vec r,t}
\right) - \bar h\left( t \right)} \right]^2 } } \right\rangle ^{{1 \mathord{\left/ {\vphantom {1 2}} \right.
\kern-\nulldelimiterspace} 2}}  = L^\alpha  g\left( {\frac{t}{{L^z }}} \right) \label{0},
\end{equation}
where $\bar h\left( t\right)$ is the mean height of the interface
at time $t$, $\alpha$ is the roughness exponent of the interface
and $z$ is the dynamic exponent that describes the scaling of the
relaxation time with $L$ - which is the size of the system. The
brackets $\left\langle \cdots  \right\rangle$ denotes noise
averaging. The scaling function $g\left(u\right)$ behaves like
$g\left(u\right) \sim u^{\beta}$ (where $\beta$ is the growth
exponent) for small $u$'s (i.e. for $t \ll L^z$) and like a
constant (i.e. $g\left(u\right) \sim const$) for large $u$'s (i.e.
for $t \gg L^z$). It is easily verified from eq. (\ref{0}) that
$\beta=\alpha/z$. The scaling exponents $\alpha$ and $z$ describe
the asymptotic behavior of the growing interface in the
hydrodynamic limit.

The KPZ equation with uncorrelated noise has been well studied.
For the one dimensional case one can easily obtain exact results
of $\alpha = {1 \mathord{\left/ {\vphantom {1 2}} \right.
\kern-\nulldelimiterspace} 2}$ and $z = {3 \mathord{\left/
{\vphantom {3 2}} \right. \kern-\nulldelimiterspace} 2}$ by
mapping the KPZ equation into the Burgers equation \cite{kpz86} or
by using the Fokker Planck equation associated with the Langevin
form given by eq. (\ref{1}) \cite{barabasi95}. However, for higher
dimensions $\left( {d > 1} \right)$ there are no exact results and
the critical exponents have been evaluated numerically or obtained
using various analytical methods (for a review see
\cite{barabasi95, Healy95}).

The noise in the KPZ equation is a result of a physical process. As such it must be correlated in space and in
time. If the correlations in space and time are short ranged it may be expected that the long distance and the
long time behavior of the system characterized by the exponents $\alpha$ and $z$ are those obtained in the case
of uncorrelated noise. There may be, however, situations in which the decay of correlations in the noise is
algebraic.

Indeed, in some experimental situations the measured scaling exponents are larger than the values predicted by
KPZ \cite{barabasi95,Healy95}. A possible explanation of such a departure from KPZ behavior may be long-range
correlations in the noise. Such experimental results serve as serve as a motivation for the study of systems
with correlated noise in spite of the fact that direct evidence for long range correlations in the noise is
usually lacking.

Many studies of growth models with noise that is algebraically correlated in space but uncorrelated in time
described by
\begin{equation}
\left\langle {\eta \left( {\vec r,t} \right)} \right\rangle  = 0
 \label{2},
\end{equation}
and
\begin{equation}
\left\langle {\eta \left( {\vec r,t} \right)\eta \left( {\vec r',t} \right)} \right\rangle  = 2D_0 \left| {\vec
r - \vec r'} \right|^{2\rho  - d} \delta \left( {t - t'} \right)
 \label{3},
\end{equation}
have been published in the last decade. These include discrete one-dimensional models (BD
\cite{Peng91,Meakin89,Amar91}, SOS \cite{Amar91,Margolina90}, and direct (discrete) integration of the KPZ
equation \cite{Peng91}). Many researchers studied the KPZ equation with such noise
\cite{medina89,Healy90,Zhang90,Hen91,katzav99,Frey99,Janssen99,chat98}) and obtained different predictions. In
spite of the differences in the predicted values of the critical exponents, a common picture seems to result
from all methods, namely: for small $\rho $'s the critical exponents are the same as for the case of
uncorrelated noise. Then, for $\rho $'s above a certain critical value $\rho _c $ the exponents become $\rho
$-dependent.

In sharp contrast to the variety of numerical results and
theoretical predictions for the critical exponents of the KPZ
equation with spatially correlated noise, only few results are
available for the KPZ equation with temporally correlated noise -
not to mention noise that is both spatially and temporally
correlated. Similar to eqs. (\ref{2})-(\ref{3}), temporally
correlated noise with zero mean can be described by
\begin{equation}
\left\langle {\eta \left( {\vec r,t} \right)\eta \left( {\vec
r',t} \right)} \right\rangle  = 2D_0 \left( {\vec r - \vec r'}
\right)\left| {t - t'} \right|^{2\phi  - 1}
\label{4},
\end{equation}
where $\phi $ characterizes the decay of the correlations over
time (it is assumed that $\phi  < {1 \mathord{\left/ {\vphantom {1
2}} \right. \kern-\nulldelimiterspace} 2}$ or otherwise the
correlations does not decay, but rather increases with time).

The first theoretical prediction of the critical exponents of KPZ in the presence of this type of noise is due
to Medina et al. [13] that used Dynamic Renormalization Group (DRG) analysis to study this problem. They solved
the DRG equations numerically in one-dimension, for the case where $D_0$ is a short range function, and found
out, just like for spatially correlated noise, that for small enough $\phi $'s the correlations are irrelevant.
They claim that for $\phi  > 0.167$ the correlations become relevant, and the roughness exponent can be fitted
numerically to
\begin{equation}
\alpha _{DRG} \left( \phi  \right) = 1.69\phi  + 0.22
\label{5}.
\end{equation}
The dynamic exponent can then be obtained using the scaling
relation
\begin{equation}
z_{DRG} \left( \phi  \right) = \frac{{2\alpha _{DRG} \left( \phi
\right) + 1}}{{1 + 2\phi }} \label{6}.
\end{equation}

These predictions have been checked numerically by Lam et al.
\cite{Lam92} using the Ballistic Deposition model. They found
sensible agreement between the DRG prediction and the numerical
values they obtained. However, substantial deviations were found,
centered around the expected threshold point $\phi _0  = 0.167$.
Thus, the authors believe that these discrepancies are due to a
crossover effect in the simulation and not due to any
approximation in the DRG calculation.

Apart from the above mentioned DRG result, there is only one more
result for KPZ with temporally correlated noise due to Ma and Ma
\cite{Ma93} who used a Flory-like Scaling Approach (SA),
originally suggested in the white-noise KPZ context, by Hentschel
and Family \cite{Hen91}. Ma and Ma obtained the following
strong-coupling roughness exponent
\begin{equation}
\alpha _{SA} \left( \phi  \right) = \frac{{2 + 4\phi }}{{2\phi  +
d + 3}}
\label{7},
\end{equation}
and the following dynamic exponent
\begin{equation}
z_{SA} \left( \phi  \right) = \frac{{2d + 4}}{{2\phi  + d + 3}}
\label{8}.
\end{equation}
These values are said to describe the strong-coupling scaling
exponents for all values of the parameter $\phi $, and for every
dimension $d$. Actually, it is easily verified that these
expression reduce to the well-know white-noise KPZ results in
one-dimension when $\phi  = 0$.

This prediction for the critical exponents is obviously different
from the previous one-dimensional DRG result in two respects.
First, Ma and Ma do not predict that for small enough $\phi $'s
the temporal correlations are irrelevant, so obviously they rule
out the threshold value of $\phi$, $\phi _0 $. Second, for $\phi
> 0.167$ the two approaches yield different numerical values for
the scaling exponents.

This situation, where only two theoretical predictions are
available for the KPZ problem in the presence of temporally
correlated noise, especially when one of them (DRG) is a
one-dimensional result, certainly calls for a clarification of
this issue. This problem is further complicated by the fact that
only one numerical study \cite{Lam92}, and only in one dimension,
is available.

At this point it is interesting to mention another result for the
KPZ equation in the presence of noise with special mixed
spatio-temporal correlations (non seperable noise correlator
$D(q,\omega)$). This is a case where in contrast to systems where
the noise is only suspected to be of long range, here long range
correlations in the noise follow from direct physical arguments.
This problem has been studied both numerically and analytically by
Li et al. \cite{Li96} with good agreement between the analytical
and numerical values. Since we deal with this problem in section
VI we will not discuss it further now.

In this paper we develop a self-consistent approach to deal with nonlinear Langevin equations, such as KPZ, with
temporally correlated noise. Actually, as will be seen in section VI, this approach can be easily generalized to
spatio-temporally correlated noise. We begin with a brief derivation of the scaling exponents of the linear
theory (also known as the Edwards-Wilkinson equation) in the presence of temporally correlated noise.Then, the
full time-dependent two-point function for the linear problem is derived. This result will serve as a reference
for the more general nonlinear discussion. In section III concepts emanating from a previous self-consistent
Fokker-Planck expansion to the KPZ equation are reviewed. In section IV the time-dependant self-consistent
approach is established. It is shown that analysis of the time-dependant self-consistent equation in the limit
of short times and long times yields two static equations that are an interesting generalization of the former
self-consistent Fokker-Planck expansion.

In section V a detailed asymptotic solution of the self-consistent equations is obtained. In this section, we
derive the different possible phases and their corresponding scaling exponents. Special attention is given to
the results in one dimension. Section VI generalizes the previous results to the case of noise with arbitrary
spatio-temporal correlations, and two elaborated examples are given. At the end, in section VII a brief summary
of the results obtained in this paper is presented.

\section{THE LINEAR THEORY - THE EDWARDS-WILKINSON EQUATION}

At the beginning of this paper we would like to discuss first the
Linear theory (i.e. the KPZ equation with its coupling constant
set to zero -$\lambda  = 0$), namely the Edwards-Wilkinson (EW)
equation \cite{EW}, with temporally correlated noise. The
Edwards-Wilkinson equation is
\begin{equation}
\frac{{\partial h}}{{\partial t}}\left( {\vec r,t} \right) = \nu
\nabla ^2 h + \eta \left( {\vec r,t} \right)
\label{9},
\end{equation}
As mentioned above, in this paper we discuss temporally correlated
noise characterized by
\begin{equation}
\left\langle {\eta \left( {\vec r,t} \right)\eta \left( {\vec
r',t'} \right)} \right\rangle  = 2D_0 \delta \left( {\vec r - \vec
r'} \right)\left| {t - t'} \right|^{2\phi  - 1}
\label{10},
\end{equation}
where the case of uncorrelated noise corresponds to the limit
$\phi  = 0$.

The interface that grows under these conditions is known to be self-affine, which means that if the spatial
coordinates are scaled by a factor of b (i.e. $\vec r \to \vec r' = b\vec r$) then if we perform the
transformations $t \to t' = b^z t$ and $h \to h' = b^\alpha  h$ (with the appropriate scaling exponents $\alpha
$-the roughness exponent, and $z$-the dynamic exponent) as well, the statistical properties of the surface are
left invariant. Since the growth equation (\ref{9}) is linear, following ref. \cite{barabasi95} it is possible
to extract the scaling exponents by scaling $\vec r,t$ and $h$ in the equation according to the above-mentioned
transformation. But first, we have to realize that under this transformation the noise term scales like $\eta
\to \eta ' = b^{{{\left( {z\left( {2\phi  - 1} \right) - d} \right)} \mathord{\left/ {\vphantom {{\left(
{z\left( {2\phi  - 1} \right) - d} \right)} 2}} \right. \kern-\nulldelimiterspace} 2}} \eta $ (see ref.
\cite{barabasi95}). Using this we can plug it back into the EW equation and we get
\begin{equation}
b^{\alpha  - z} \frac{{\partial h'}}{{\partial t'}}\left( {\vec r',t'} \right) = b^{\alpha  - 2} \nu \nabla '^2
h' + b^{{{\left( {z\left( {2\phi  - 1} \right) - d} \right)} \mathord{\left/
 {\vphantom {{\left( {z\left( {2\phi  - 1} \right) - d} \right)} 2}} \right.
 \kern-\nulldelimiterspace} 2}} \eta '\left( {\vec r',t'} \right)
\label{11}
\end{equation}
Now, imposing the requirement that eq. (\ref{9}) remains invariant under this scaling transformation, namely
requiring that both equations (eqs. (\ref{9}) and (\ref{11})) should be exactly the same, we get
\begin{equation}
z = 2\quad \quad and\quad \quad \alpha  = {{\left( {4\phi  + 2-d} \right)} \mathord{\left/
 {\vphantom {{\left( {4\phi  + 2-d} \right)} 2}} \right.
 \kern-\nulldelimiterspace} 2}
\label{12}.
\end{equation}
(this gives the roughness exponent as long as the resulting $\alpha$ is positive, otherwise the surface is
flat). It is easily seen that this result reduces to the standard EW exponents (i.e. for the EW equation with
uncorrelated noise) in the limit of $\phi  = 0$.

This simple result shows that temporally correlated noise tends to
make the surface rougher (a bigger roughness exponent $\alpha $
implies a rougher surface).

\begin{figure}[htb]
\includegraphics[width=8cm]{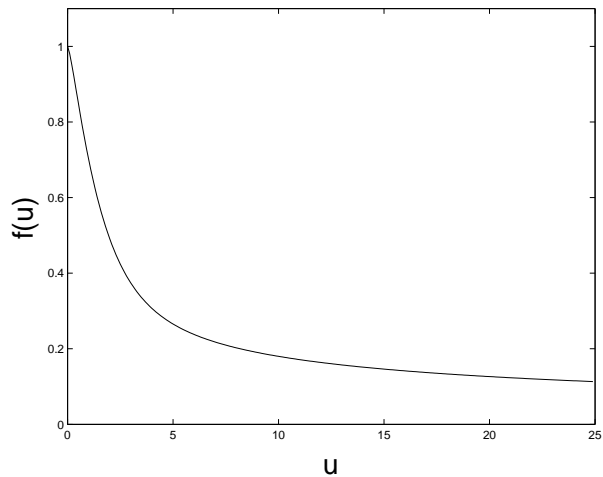}
\caption{The scaling function $f(u)$ ($\phi=1/4$ was taken for this illustration). One can see an
exponential-like decay for small u's, and a power law decay for large u's.} \label{scaling}
\end{figure}

The information extracted so far regrading the EW equation in the presence of temporally correlated noise could
have been satisfactory. However, because we are interested in obtaining the exponents of the nonlinear theory as
well, we would like to gain as much insight into the behavior of the linear problem, so that it might help us
when dealing with the KPZ nonlinearity. For example, because of the linear character of eq. (\ref{9}), we can
obtain the scaling form (and recover the exponents) by solving the growth equation exactly. Fourier transforming
eq. (\ref{9}) in space and time we obtain
\begin{equation}
h_{q\omega }  = \frac{{\eta _{q\omega } }}{{i\omega  + \nu q^2 }}
\label{12.1},
\end{equation}
where $\eta_{q\omega }$ is the Fourier transform of $\eta \left(
{\vec r,t} \right)$. Thus, using the Fourier transform of eq.
(\ref{10}), we obtain the dynamical structure factor (or the two
point correlation function)
\begin{equation}
\Phi_{q\omega}=\left\langle {h_{q\omega } h_{ - q, - \omega } } \right\rangle  = 2D_0 \frac{{\omega ^{ - 2\phi }
}}{{\omega ^2 + \nu ^2 q^4 }} \label{12.2}.
\end{equation}
By Fourier transforming back we get
\begin{equation}
\Phi _q \left( t \right) = \left\langle {h_q \left( 0 \right)h_{ - q} \left( t \right)} \right\rangle  =
\frac{{D_0 }}{{\nu ^{1 + 2\phi } \cos \left( {\pi \phi } \right)}}q^{ - 2 - 4\phi } f\left( {\nu q^2 t} \right)
\label{12.3}.
\end{equation}
Here $f(u)$ is a scaling function that can be written explicitly as
\begin{equation}
f\left( u \right) = \frac{{\cos \left( {\pi \phi } \right)}}{\pi }\int\limits_{ - \infty }^\infty {\frac{{y^{ -
2\phi } }}{{y^2  + 1}}e^{iyu} dy}  = \cosh \left( u \right) - \frac{{u^{1 + 2\phi } }}{{\Gamma \left( {2 + 2\phi
} \right)}}{}_1F_2 \left( {\left. {\begin{array}{*{20}c}
   1  \\
   {\phi  + 1,\phi  + {\textstyle{3 \over 2}}}  \\
\end{array}} \right|\frac{{u^2 }}{4}} \right)
\label{12.4}.
\end{equation}
where $\Gamma(x)$ is just Euler's Gamma function, and $_1F_2$ is a generalized hypergeometric function. The
function $f(u)$ is also plotted in Fig. \ref{scaling}.
As can be seen in the figure, the scaling function behaves like a constant for small $u$'s (this corresponds to
short times, that is for $\nu q^2t \ll 1$). At the other extreme, i.e. for large $u$'s, this function decays
algebraically. In order to be sure of this power law tail, and to obtain its exact shape we calculated the
leading behaviors for small and large $u$'s and obtained
\begin{equation}
f\left( u \right) \sim \left\{ \begin{array}{l}
 1 - \frac{1}{{\Gamma \left( {2\phi  + 2} \right)}}u^{1 + 2\phi }  +  \cdots \quad \quad \quad \quad \quad \quad \quad u \ll 1 \\
 \frac{{u^{2\phi  - 1} }}{{\Gamma \left( {2\phi } \right)}}\left( {1 + \frac{{4\left( {1 - \phi } \right)\left( {{\textstyle{1 \over 2}} - \phi } \right)}}{{u^2 }} +  \cdots } \right)\quad \quad \quad \quad u \gg 1 \\
 \end{array} \right.
\label{12.5},
\end{equation}
so that $f(u) \sim u^{-(1-2\phi)}$ for large $u$'s.

Naturally, the scaling exponents can be recovered easily from $\Phi_q(t)$. Since $\Phi_q(t)$ depends on time
only through the combination $\nu q^2t$ we identify the dynamic exponent as the power of $q$ in this scaling
form, so that here $z=2$. In addition, it can be seen that for small $q$'s, $\Phi_q(t) \sim q^{-2-4\phi}$. Thus
we identify the exponent $\Gamma=2+4\phi$ that can be translated into the roughness exponent via the relation
$\alpha=(\Gamma-d)/2$ (see eq. (\ref{36}) below), so that we recover $\alpha=(4\phi+2-d)/2$.

The results obtained in this section will serve us later. First,
it might be interesting to compare these results with the results
obtained for the nonlinear theory (for example, in the
weak-coupling regime of the KPZ equation). Second, we will use
the scaling function of the linear theory as an ansatz for the
integral equation that will determine the scaling exponents of
the strong-coupling phase of the nonlinear theory.

\section{NONLINEAR THEORY - THE KPZ EQUATION}

We proceed now to the much harder nonlinear case that poses many
technical difficulties already in the uncorrelated case.

The method we present in the following section is based on the
same general ideas as the self consistent expansion (SCE) used for
systems with noise that is uncorrelated in time
\cite{SE98,katzav99,SE92}. Namely, an expansion around an optimal
linear system. The SCE is based on constructing a Fokker-Planck
equation for the probability distribution of the height function.
This step is based on the fact that the noise is not correlated
in time. The self consistent expansion is formulated in terms of
the steady state structure factor (or two-point function),
$\phi_q=\left\langle {h_q h_{ - q}} \right\rangle_S$ and its
corresponding steady state decay rate that describes the rate of
decay of a disturbance of wave vector $\vec q$ in steady state
\begin{equation}
\omega _q^{-1}  = \frac{{\int_0^\infty  {\left\langle {h_q \left(
t \right)h_{ - q} \left( 0 \right)} \right\rangle dt}
}}{{\left\langle {h_q h_{ - q} } \right\rangle _S }}
\label{13}.
\end{equation}

The linear model around which the expansion is constructed is
chosen to yield the (unknown) $\phi_q$ and $\omega_q$ that appear
in it as parameters. An evaluation of $\phi_q$ and $\omega_q$ as
an expansion around that linear model leads to the coupled
equations

\begin{equation}
\phi _q  = \phi _q  + c_q \left\{ {\phi _p ,\omega _p } \right\}
\label{14},
\end{equation}
and
\begin{equation}
\omega _q  = \omega _q  + d_q \left\{ {\phi _p ,\omega _p }
\right\}
\label{15}.
\end{equation}
Within this framework, the structure factor and decay rate are
obtained by solving the coupled non-linear integral equations $c_q
\left\{ {\phi _p ,\omega _p } \right\} = 0$ and $d_q \left\{ {\phi
_p ,\omega _p } \right\} = 0$. In contrast to other expansions,
the full correction, in a given order of the expansion, of the
relevant physical quantities, is really small. In fact, it is
chosen to be zero.

\section{DERIVATION OF THE TIME DEPENDENT SELF-CONSISTENT APPROACH}

In this work (following ref. \cite{SE02}) we obtain the dynamical
structure factor $\Phi _q (t) = \left\langle {h_q \left( 0
\right)h_{ - q} \left( t \right)} \right\rangle _s$, using the
same idea of a self consistent expansion. Here too, the average
$\left\langle  \cdots  \right\rangle _s$ denotes steady-state
averaging, where $h_q \left( 0 \right)$ is measured in steady
state at time $t = 0$ and then $h_{ - q} \left( t \right)$ is
measured at some later time $t$ (also in steady state). The
dynamical structure factor $\Phi _q (t)$ normalized by $\phi _q =
\Phi _q \left( 0 \right)$ (i.e. the static structure factor) is
thus a measure of the persistence in steady state of disturbances
with wave vector $\vec q$. Because the noise is correlated in
time, we cannot use the Fokker-Planck approach, but as seen in
refs. \cite{SE02,SE02b} such an approach lends itself as an
alternative to the Fokker-Planck approach even when it is
available.

Our starting point is the field equation for $h_{q\omega }$ (the
Fourier transform in time and space of $h\left( {\vec r,t}
\right)$) obtained by Fourier transforming eq. (\ref{1})
\begin{equation}
i\omega h_{q\omega }  + \nu _q h_{q\omega }  + \sum\limits_{\ell
,\sigma ,m,\tau } {C_{q\ell m} h_{\ell \sigma } h_{m\tau } }  =
\eta _{q\omega }
\label{16},
\end{equation}
where $\nu_q=\nu q^2$, $C_{q\ell m}  = \frac{1}{{\sqrt T
}}\frac{1}{{\sqrt \Omega }}\vec \ell  \cdot \vec m\delta
_{q,\ell  + m} \delta _{\omega ,\sigma  + \tau }$, $T$ being an
assumed periodicity in time to be taken eventually to infinity,
$\Omega $ is the volume of the system (to be taken to infinity as
well) and the noise correlations are $\left\langle {\eta
_{q\omega } \eta _{ - q - \omega } } \right\rangle  = 2D^0 \left(
q \right)\omega ^{ - 2\phi }$. (Note, that in the $d+1$
dimensional space (including time), the noise is quenched
disorder!). In the Chapman-Enskog spirit (as done in refs.
\cite{Edwards69,SE02}) the equation is written in the form
\begin{equation}
\left[ {\left( {i\omega  + \omega _q } \right)h_{q\omega }  - \eta
_{q\omega }^0 } \right] + \lambda \left[ {\sum\limits_{\ell
,\sigma ,m,\tau } {C_{q\ell m} h_{\ell \sigma } h_{m\tau } }  -
\eta _{q\omega }^1 } \right] + \lambda ^2 \left[ {\left( {\nu _q -
\omega _q } \right)h_{q\omega } } \right] = 0
\label{17},
\end{equation}
where $\lambda$ is going to be taken as $1$ but is used at present
as an indicator to show the construction of the perturbation
expansion as an expansion in $\lambda$. The noise is split into
two terms $\eta_{q\omega}=\eta _{q\omega }^0 + \eta _{q\omega }^1$
such that $\left\langle {\eta _{q\omega }^0 \eta _{ - q - \omega
}^0 } \right\rangle  = D_{q\omega }$ and the correct $\Phi
_{q\omega }$ (i.e. the Fourier transform in time of the "dynamical
structure factor" $\Phi _q (t)$) is given by $\Phi _{q\omega }  =
\frac{{D_{q\omega } }}{{\omega ^2  + \omega_q ^2 }}$. This choice
implies that ignoring the $\lambda$ and $\lambda ^2$ terms in eq.
(\ref{17}), we still obtain from a linear equation the correct
$\Phi _{q\omega }$. In contrast to the case of short range
correlated noise where $\omega_q$ is defined by eq. (\ref{13}),
we must employ here a more general definition. The reason is that
the power law found to describe the tail of $\Phi_q(t)$ for long
times renders the expression on the right hand side of eq.
(\ref{13}) infinite. Therefore, our definition of $\omega_q$ is
based on the assumption of a scaling form of $\Phi_q(t)$ - namely
\begin{equation}
\Phi_q(t)=\phi_q f(\omega_qt)
\label{17.5},
\end{equation}
(it can be easily verified that the dynamical structure factor of
the linear theory given by eq. (\ref{12.3}) indeed obeys this
scaling law). The "decay rate", $\omega_q$, is defined as that
parameter that will make eq. (\ref{17.5}) a good approximation
for small $q$'s and over the whole time range.

Now, Eq. (\ref{17}) enables to obtain $h_{q\omega}$ explicitly to
second order in $\lambda$. The expression for $h_{q\omega }$ is
multiplied into its complex conjugate and only terms up to second
order in $\lambda $ are retained. At the end the expressions are
averaged over $\eta _{q\omega }$ and we get
\begin{eqnarray}
 \left( {\omega ^2  + \omega _q^2 } \right)\Phi _{q\omega }  &=& D_{q\omega }  + 2\lambda ^2 \sum\limits_{\ell ,m,\sigma ,\tau } {C_{q\ell m} C_{q - \ell  - m} \Phi _{\ell \sigma } \Phi _{m\tau } }    \nonumber\\
  &+& \lambda ^2 \left( {2D^0 \left( q \right)\omega ^{ - 2\phi }  - D_{q\omega } } \right) - 2\lambda ^2 \left( {\nu _q  - \omega _q } \right)\omega _q \Phi _{q\omega }    \nonumber\\
  &+& 4\lambda ^2 \sum\limits_{\ell ,m,\sigma ,\tau } {C_{q\ell m} C_{\ell q - m} \Phi _{q\omega } \Phi _{m\tau } \frac{{ - i\omega  + \omega _q }}{{i\sigma  + \omega _\ell  }}} \nonumber\\
  &+& 4\lambda ^2 \sum\limits_{\ell ,m,\sigma ,\tau } {C_{q - \ell  - m} C_{\ell  - qm} \Phi _{m\tau } \Phi _{q\omega } \frac{{i\omega  + \omega _q }}{{ - i\sigma  + \omega _\ell  }}}
\label{18}.
\end{eqnarray}

Now $\lambda $ is set to be $1$. The result is an equation of the
form $\Phi _{q\omega }  = \Phi _{q\omega }  + e\left\{ {\Phi
_{\ell \sigma } } \right\}$. Equating $e\left\{ {\Phi _{\ell
\sigma } } \right\}$ to zero yields
\begin{eqnarray}
 \left[ {\omega ^2  + \omega _q^2  + 2\left( {\nu _q  - \omega _q } \right)\omega _q } \right]\Phi _{q\omega }  - 2\sum\limits_{\ell ,\sigma ,m,\tau } {\left| {C_{q\ell m} } \right|^2 \Phi _{\ell \sigma } \Phi _{m\tau }} &+& \nonumber\\
 + 4\sum\limits_{\ell ,\sigma ,m,\tau } {C_{q\ell m} C_{\ell qm} \Phi _{q\omega } \Phi _{m\tau } \left[ {\frac{{ - i\omega  + \omega _q }}{{i\sigma  + \omega _\ell  }} + \frac{{i\omega  + \omega _q }}{{ - i\sigma  + \omega _\ell  }}} \right]} &=& 2D^0 \left( q \right)\omega ^{ - 2\phi }
\label{19}.
\end{eqnarray}
We divide the last equation by $\left( {\omega ^2  + \omega _q^2
} \right)$ and using the definition  $C_{q\ell m}  =
\frac{{A_{\ell ,q - \ell } \delta _{q,\ell  + m} }}{{\sqrt
{\Omega T} }}$ (as well as letting $\Omega$ and $T$ tend to
infinity) we obtain
\begin{eqnarray}
&& \left[ {1 + 2\frac{{\left( {\nu _q  - \omega _q } \right)\omega _q }}{{\omega ^2  + \omega _q^2 }}} \right]\Phi _{q\omega }  - 2\int {\frac{{d^d \ell }}{{\left( {2\pi } \right)^d }}\frac{{d\sigma }}{{2\pi }}\frac{{\left| {A_{\ell ,q - \ell } } \right|^2 \Phi _{\ell \sigma } \Phi _{q - \ell ,\omega  - \sigma } }}{{\omega ^2  + \omega _q^2 }}}  \nonumber\\
&+& 4\int {\frac{{d^d \ell }}{{\left( {2\pi } \right)^d }}\frac{{d\sigma }}{{2\pi }}A_{\ell ,q - \ell } A_{q,q - \ell } \left[ {\frac{{\Phi _{q\omega } \Phi _{q - \ell ,\omega  - \sigma } }}{{\left[ {i\omega  + \omega _q } \right]\left[ {i\sigma  + \omega _\ell  } \right]}} + \frac{{\Phi _{q\omega } \Phi _{q - \ell ,\omega  - \sigma } }}{{\left[ { - i\sigma  + \omega _\ell  } \right]\left[ { - i\omega  + \omega _q } \right]}}} \right]}  \nonumber\\
&=& \frac{{\omega ^{ - 2\phi } }}{{\omega ^2  + \omega _q^2 }}2D^0 \left( q \right) \label{20}.
\end{eqnarray}

The last equation is the basic equation for our following
discussion. We consider first the small $\omega$ behavior (more
specifically $\omega/\omega_q \ll 1$) that corresponds to the
long time decay of the time dependent structure factor.

\subsection{Long time decay of the structure factor}
The first small $\omega$ simplification is obtained by neglecting $\omega/\omega_q$. This yields
\begin{eqnarray}
&& \left[ {1 + 2\frac{{\left( {\nu _q  - \omega _q } \right)\omega _q }}{{\omega _q^2 }}} \right]\Phi _{q\omega }  - 2\frac{1}{{\omega _q^2 }}\int {\frac{{d^d \ell }}{{\left( {2\pi } \right)^d }}\frac{{d\sigma }}{{2\pi }}\left| {A_{\ell ,q - \ell } } \right|^2 \Phi _{\ell \sigma } \Phi _{q - \ell ,\omega  - \sigma } }  \nonumber\\
&& \quad \quad \quad  + 8\frac{{\Phi _{q\omega } }}{{\omega _q
}}\int {\frac{{d^d \ell }}{{\left( {2\pi } \right)^d
}}\frac{{d\sigma }}{{2\pi }}A_{\ell ,q - \ell } A_{q,q - \ell }
\frac{{\omega _\ell  }}{{\sigma ^2  + \omega _\ell ^2 }}\Phi _{q
- \ell ,\omega  - \sigma } }  = \frac{{2D^0 \left( q
\right)}}{{\omega _q^2 }}\omega ^{ - 2\phi }
\label{21}.
\end{eqnarray}
Fourier transforming back from frequency domain $\omega$ to real
time $t$, we obtain
\begin{eqnarray}
 \left[ {1 + 2\frac{{\left( {\nu _q  - \omega _q } \right)\omega _q }}{{\omega _q^2 }}} \right]\Phi _q \left( t \right) &-& \frac{2}{{\omega _q^2 }}\int {\frac{{d^d \ell }}{{\left( {2\pi } \right)^d }}\left| {A_{\ell ,q - \ell } } \right|^2 \Phi _\ell  \left( t \right)\Phi _{q - \ell } \left( t \right)}  \nonumber\\
&+& \frac{8}{{\omega _q }}\int {\frac{d^d \ell}{{\left( {2\pi } \right)^d }}  A_{\ell ,q - \ell } A_{q,\ell  - q} \int_{ - \infty }^\infty  {dt'} e^{ - \omega _\ell  \left| {t'} \right|} \Phi _{q - \ell } \left( {t'} \right)\Phi _q \left( {t - t'} \right)}  \nonumber \\
&=& \frac{{D^0 \left( q \right)}}{{\Gamma \left( {2\phi }
\right)\cos \left( {\pi \phi } \right)\omega _q^2 }}t^{ - \left(
{1 - 2\phi } \right)}
\label{22},
\end{eqnarray}
where on the right hand side we have written only the leading
large-$t$ behavior.

This result suggests that in the long-time limit, the
time-dependent two-point function has an algebraic decay of the
general form
\begin{equation}
\Phi _q \left( t \right)\sim A^{\infty}  \phi _q \left( {\omega _q
t} \right)^{ - \gamma }
\label{23},
\end{equation}
where $A^\infty$ is a numerical constant, $\phi _q $ is the
steady-state two-point function, and $\gamma $ is an exponent that
will be determined later.

Equipped with the last result we can see that the first integral on the left hand side of eq. (\ref{22}) is
negligible compared to the other terms on that side in the long-time limit. The reason is that this integral
decays as $t^{-2\gamma}$, while the other terms decay as $t^{-\gamma}$, making that integral subdominant for
large $t$'s.

Next, using this simplification as well as the scaling form
(\ref{17.5}), we analyze eq. (\ref{22}) for small $q$'s (i.e. in
the large scale limit) in the spirit of refs.
\cite{SE98,katzav99}. In order to achieve that, we break up the
integral into the sum of two contributions corresponding to
domains of $\vec \ell$ integration, with high and low momentum.
When performing this under the assumption of long-times (i.e.
$\omega _q t \gg 1$) we obtain the following equation
\begin{eqnarray}
\frac{{2\nu _q  - \omega _q }}{{\omega _q }}\phi _q f\left(
{\omega _q t} \right)
&-& \frac{8}{{\left( {2\pi } \right)^d }}\frac{{\phi _q }}{{\omega _q }}f\left( {\omega _q t} \right)\left[ {\hat D_1 q^2  + \int_{}^{q_0 } {d^d \ell A_{\ell ,q - \ell } A_{q,\ell  - q} \frac{{\phi _{q - \ell } }}{{\omega _{q - \ell } }}F_1 \left( {\left\{ f \right\},{\textstyle{{\omega _\ell  } \over {\omega _{q - \ell } }}}} \right)} } \right] \nonumber\\
&=& \frac{{D^0 \left( q \right)\left( {\omega _q } \right)^{ - 1
- 2\phi } }}{{A^\infty \Gamma \left( {2\phi } \right)\cos \left(
{\pi \phi } \right)}}\left( {\omega _q t} \right)^{2\phi - 1}
\label{24}.
\end{eqnarray}
where $f$, inside the curly brackets, is just the scaling
function, $q_0$ is the upper cut-off of the small $\left| {\vec
\ell } \right|$ region, and $\hat D_1$ is a constant that comes
from the contribution of the large $\left| {\vec \ell } \right|$
region of the first integral (see ref. \cite{SE98} section VI,
where such an estimation of the contribution of large momenta is
also employed). In addition, we used the following notation for
the integral $F_1$
\begin{equation}
F_1 \left( {\left\{ f \right\},{\textstyle{{\omega _\ell  } \over
{\omega _{q - \ell } }}}} \right) = \int_0^\infty  {e^{ -
{\textstyle{{\omega _\ell  } \over {\omega _{q - \ell } }}}x}
f\left( x \right)dx}
\label{F1}.
\end{equation}

We conclude that $\Phi _q \left( t \right)\sim A^\infty  \phi _q
\left( {\omega _q t} \right)^{ - \left( {1 - 2\phi } \right)}$
(i.e. $\gamma  = 1 - 2\phi $). And eq. (\ref{24}) can be
re-written as a time-independent equation relating the $\phi$'s
and the $\omega$'s
\begin{eqnarray}
&&A^\infty  \frac{{\phi _q }}{{\omega _q }}\left[ {2\nu _q  +D_1q^2- \omega _q  + \frac{8}{{\left( {2\pi }
\right)^d }}\int_{}^{q_0 } {d^d \ell A_{\ell ,q - \ell } A_{q,\ell  - q} \frac{{\phi _{q - \ell } }}{{\omega _{q
- \ell } }}F_1 \left( {\left\{ f \right\},{\textstyle{{\omega _\ell  } \over {\omega _{q - \ell }
}}}} \right)} } \right] \nonumber\\
&&\quad\quad=\frac{{D^0 \left( q \right)\left( {\omega _q } \right)^{ - 1 - 2\phi } }}{{\Gamma \left( {2\phi }
\right)\cos \left( {\pi \phi } \right)}}
\label{25}.
\end{eqnarray}

It is interesting to compare the result above for $\Phi_q (t)$ with the decay in the case where the noise is not
correlated in time (namely, when $\phi=0$) \cite{SE02}. In that case the long time behavior of $\Phi_q (t)$ is
given by
\begin{equation}
\Phi _q \left( t \right) \propto \left( {\omega _q t}
\right)^{\frac{{d - 1}}{{2z}}} \exp \left[ { - \left( {\omega _q
t} \right)^{\frac{1}{2}} } \right]
\label{26},
\end{equation}
(i.e. a stretched exponential).

The limit as $\phi$ tends to zero of $\Phi_q (t)$ should yield a
short range decay. The expression in eq. (\ref{23}) (given that
$\gamma=2\phi-1$) tends to a function that scales as $t^{-1}$.
This should be viewed as a function that scales as $\delta(t)$ at
large $t$'s, or a short range function. Actually, a direct
inspection on the right hand side of eq. (\ref{22}) recovers this.
Since the denominator of the right hand side contains the $\Gamma$
function, the whole expression vanish as $\phi$ tends to zero.
Checking more carefully, for $\phi=0$ the right hand side of eq.
(\ref{22}) is proportional to $\exp \left[ { - \omega _q t}
\right]$. If we try now a solution $\Phi_q(t) \propto \exp \left[
{ - \omega _q t} \right]$ we find that it does not work. The
reason for that and how to obtain the correct asymptotic behavior
(eq. {\ref{26})) is detailed in ref. \cite{SE02}.

To complete the picture given so far, that is after finding the
power law that governs the of the structure factor, we need to
know the steady state structure factor $\phi_q$ and its associated
"decay rate" $\omega_q$ that will characterize the short time
decay as well.

\subsection{Steady state properties}
In this part we obtain another equation, that together with eq.
(\ref{25}), forms a complete set of coupled equations that will
yield the small $q$ dependence of $\phi_q$ and $\omega_q$. In
order to achieve this we would like to discuss eq. (\ref{20}) in
the limit of short-times as well. Here, it is more convenient to
treat the time-dependent equation directly, so we Fourier
transform eq. (\ref{20}) to yield
\begin{eqnarray}
 \Phi _q \left( t \right) &+& \left( {\nu _q  - \omega _q } \right)\int\limits_{ - \infty }^\infty  {dt'e^{ - \omega _q \left| {t'} \right|} \Phi _q \left( {t - t'} \right)}  -  \nonumber\\
 &-& \frac{2}{{\left( {2\pi } \right)^d }}\int {d^d \ell \left| {A_{\ell ,q - \ell } } \right|^2 \int\limits_{ - \infty }^\infty  {dt'\frac{{e^{ - \omega _q \left| {t'} \right|} }}{{2\omega _q }}\Phi _{q - \ell } \left( {t - t'} \right)\Phi _\ell  \left( {t - t'} \right)} }  +  \nonumber\\
 &+& \frac{4}{{\left( {2\pi } \right)^d }}\int {d^d \ell A_{\ell ,q - \ell } A_{q,q - \ell } \left\{ {\int\limits_0^\infty  {dt'\int\limits_0^\infty  {dt''e^{ - \omega _\ell  t' - \omega _q t''} \Phi _{q - \ell } \left( {t'} \right)\Phi _q \left( {t - t' - t''} \right)} }  + } \right.}  \nonumber\\
 &+&\left. { \int\limits_{ - \infty }^0 {dt'\int\limits_{ - \infty }^0 {dt''e^{\omega _\ell  t' + \omega _q t''} \Phi _{q - \ell } \left( {t'} \right)\Phi _q \left( {t - t' - t''} \right)} } } \right\} = \frac{{D^0 \left( q \right)}}{\pi }\int\limits_{ - \infty }^\infty  {\frac{{\omega ^{ - 2\phi } e^{i\omega t} }}{{\omega ^2  + \omega _q^2 }}d\omega }
\label{27}.
\end{eqnarray}

Setting $t=0$ and following the same steps described above for
long-times (i.e. breaking the $\vec \ell $-integration into large
and small $\left| {\vec \ell } \right|$ regions, and discussing
the small q behavior of each) gives the following short-time
evaluation of eq. (\ref{27})
\begin{eqnarray}
 \nu _q \phi _q  &+& \frac{8}{{\left( {2\pi } \right)^d }}\phi _q \left[ {\hat E_1 q^2  + \int_{}^{q_0 } {d^d \ell \frac{{A_{\ell ,q - \ell } A_{q,\ell  - q} }}{{\omega _\ell  }}\phi _{q - \ell } F_2 \left( {\left\{ f \right\},{\textstyle{{\omega _{q - \ell } } \over {\omega _\ell  }}},{\textstyle{{\omega _q } \over {\omega _\ell  }}}} \right)} } \right] \nonumber\\
&-& \frac{2}{{\left( {2\pi } \right)^d }}\left[ {\frac{1}{{\omega
_q }}\int_{}^{q_0 } {d^d \ell \left| {A_{\ell ,q - \ell } }
\right|^2 \phi _{q - \ell } \phi _\ell  F_3 \left( {\left\{ f
\right\},{\textstyle{{\omega _{q - \ell } } \over {\omega _q
}}},{\textstyle{{\omega _\ell  } \over {\omega _q }}}} \right)} +
\hat E_2  + \frac{{\hat E_3 }}{{\omega _q^{4\phi  - 1} }}}
\right] \nonumber\\
&=& \frac{{D^0 \left( q \right)}}{{\cos \left( {\pi \phi }
\right)}}\left( {\omega _q } \right)^{ - 2\phi }
\label{28},
\end{eqnarray}
where as before $\hat E_1$, $\hat E_2$ and $\hat E_3$ are (renormalization) constants. In addition, we used the
following notations
\begin{equation}
F_2 \left( {\left\{ f \right\},{\textstyle{{\omega _{q - \ell } }
\over {\omega _\ell  }}},{\textstyle{{\omega _q } \over {\omega
_\ell  }}}} \right) = \int\limits_0^\infty
{dx\int\limits_0^\infty  {dye^{ - x - y} f\left(
{{\textstyle{{\omega _{q - \ell } } \over {\omega _\ell  }}}x}
\right)f\left( {{\textstyle{{\omega _q } \over {\omega _\ell
}}}x + y} \right)} }
\label{F2},
\end{equation}
and
\begin{equation}
F_3 \left( {\left\{ f \right\},{\textstyle{{\omega _{q - \ell } }
\over {\omega _q }}},{\textstyle{{\omega _\ell  } \over {\omega
_q }}}} \right) = \int\limits_0^\infty  {e^{ - x} f\left(
{{\textstyle{{\omega _{q - \ell } } \over {\omega _q }}}x}
\right)f\left( {{\textstyle{{\omega _\ell  } \over {\omega _q
}}}x} \right)dx}
\label{F3}.
\end{equation}

Up to this point we obtained two coupled equations for $\phi _q $
and $\omega _q $. Note that the equations above depend on the
(unknown) functional form of the scaling function $f$. We will
proceed now as far as possible without specifying that form, to
obtain many general results about the exponents. In the actual
numerical calculation of the exponents we will resort to an
approximate form of $f$ to be described later.

We would like now to solve eqs. (\ref{25}) and (\ref{28}) in the
limit of small $q$'s. For convenience we rewrite these equations
using the explicit form of $A_{\ell,m}$ as
\begin{equation}
\phi _q \left[ {\left( {2\nu  + D_1 } \right)q^2  - \omega _q  +
J^ <  \left( q \right)} \right] = \frac{{D^0 \left( q
\right)\left( {\omega _q } \right)^{ - 2\phi } }}{{A^\infty  \cos
\left( {\pi \phi } \right)\Gamma \left( {2\phi } \right)}}
\label{29},
\end{equation}
and
\begin{equation}
\left( \nu  + E_1 \right)q^2 \phi _q + I_1^ < \left( q \right)\phi _q  - I_2^ <  \left( q \right) - E_2 -
\frac{{E_3 }}{{\omega _q^{4\phi  - 1} }} = \frac{{D^0 \left( q \right)\left( {\omega _q } \right)^{ - 2\phi }
}}{{\cos \left( {\pi \phi } \right)}} \label{30}.
\end{equation}
where
\begin{equation}
J^ <  \left( q \right) = \frac{{8\lambda ^2 }}{{\left( {2\pi }
\right)^d }}\int_{}^{q_0 } {d^d \ell \frac{{\left[ {\vec \ell
\cdot \left( {\vec q - \vec \ell } \right)} \right]\left[ {\vec q
\cdot \left( {\vec q - \vec \ell } \right)} \right]}}{{\omega _{q
- \ell } }}\phi _{q - \ell } F_1 \left( {\left\{ f
\right\},{\textstyle{{\omega _\ell  } \over {\omega _{q - \ell }
}}}} \right)}
\label{31},
\end{equation}
\begin{equation}
I_1^ <  \left( q \right) = \frac{{8\lambda ^2 }}{{\left( {2\pi }
\right)^d }}\int_{}^{q_0 } {d^d \ell \frac{{\left[ {\vec \ell
\cdot \left( {\vec q - \vec \ell } \right)} \right]\left[ {\vec q
\cdot \left( {\vec q - \vec \ell } \right)} \right]}}{{\omega
_\ell  }}\phi _{q - \ell } F_2 \left( {\left\{ f
\right\},{\textstyle{{\omega _{q - \ell } } \over {\omega _\ell
}}},{\textstyle{{\omega _q } \over {\omega _\ell  }}}} \right)}
\label{32},
\end{equation}
and
\begin{equation}
I_2^ <  \left( q \right) = \frac{{2\lambda ^2 }}{{\left( {2\pi }
\right)^d }}\int_{}^{q_0 } {d^d \ell \frac{{\left[ {\vec \ell
\cdot \left( {\vec q - \vec \ell } \right)} \right]^2 }}{{\omega
_q }}\phi _{q - \ell } \phi _\ell  F_3 \left( {\left\{ f
\right\},{\textstyle{{\omega _{q - \ell } } \over {\omega _q
}}},{\textstyle{{\omega _\ell  } \over {\omega _q }}}} \right)}
\label{33}.
\end{equation}

Notice that the integrals in eqs. (\ref{31})-(\ref{33}) are cut by
$q_0$. $q_0$ is chosen in such a way that below it $\phi _q$ and
$\omega _q $ are expected to be power laws in $q$,
\begin{equation}
\phi _q  = Aq^{ - \Gamma }
\label{34},
\end{equation}
and
\begin{equation}
\omega _q  = Bq^z
\label{35},
\end{equation}
where $z$ is the dynamic exponent, and $\Gamma$ is related to the
roughness exponent $\alpha $ by
\begin{equation}
\alpha  = {{\left( {\Gamma  - d} \right)} \mathord{\left/
{\vphantom {{\left( {\Gamma  - d} \right)} 2}} \right.
\kern-\nulldelimiterspace} 2}
\label{36}.
\end{equation}

As mentioned above, the integrals in eqs. (\ref{31})-(\ref{33})
are cut by $q_0 $, and therefore we can readily use these
power-laws inside the integrals. Using the power laws we can also
rewrite eqs. (\ref{29})-(\ref{30}) as
\begin{equation}
Aq^{ - \Gamma } \left[ {\left( {2\nu  + D_1 } \right)q^2  - Bq^z
+ J^ <  \left( q \right)} \right] = \frac{{D^0 \left( q
\right)\left( {Bq^z } \right)^{ - 2\phi } }}{{A^\infty  \cos
\left( {\pi \phi } \right)\Gamma \left( {2\phi } \right)}}
\label{37},
\end{equation}
and
\begin{equation}
A\left( \nu  + E_1 \right)q^{2 - \Gamma } + I_1^ <  \left( q \right)Aq^{ - \Gamma }  - I_2^ < \left( q \right) -
E_2  = \frac{{D^0 \left( q \right)\left( {Bq^z } \right)^{ - 2\phi } }}{{\cos \left( {\pi \phi } \right)}}
\label{38},
\end{equation}
where we have neglected the $E_3$ term in eq. (\ref{30}) as it is negligible compared to the left hand side in
the limit of small $q$'s (since $\phi<1/2$).

It is interesting to notice that these equations are a nontrivial
generalization of the Self-Consistent Expansion (SCE) developed in
refs \cite{SE98,SE92}. More specifically, if we take the limit of
$\phi \rightarrow 0$, and plug in $f(u)=e^{-u}$, which is the
scaling function of the linear theory when $\phi=0$, both
equations (i.e. eqs. (\ref{37})-(\ref{38})) reduce to the
equations obtained using SCE. It is a surprise to find this
similarity because the self-consistent expansion was originally
derived using the Fokker-Planck equation associated with the
Langevin-like KPZ equation, while the derivation given here deals
directly with the Langevin form. Once we realized this surprising
similarity, it is only natural to follow the asymptotic solution
that is used in the well-established SCE literature, and is
detailed for example in ref. \cite{SE98}.

\section{DETAILED ASYMPTOTIC SOLUTION}

As mentioned above, in performing the asymptotic solution of the
self-consistent equations, we follow previous work. We also focus
here, for simplicity, on the case of noise without spatial
correlations (i.e. $D^0 \left( q \right) = D_0 $). However, eqs.
(\ref{37})-(\ref{38}) are valid for any spatial correlations of
the noise (i.e. for any $D^0 \left( q \right)$), so that the more
general case is postponed to the next section.

The fist step in the asymptotic solution is to evaluate the
integrals $I_1^ <  \left( q \right)$, $I_2^ <  \left( q \right)$
and $J^ <  \left( q \right)$ using the power laws given in eqs.
(\ref{34})-(\ref{35})
\begin{equation}
I_1^ <  \left( q \right),J^ <  \left( q \right) \propto \left\{
\begin{array}{l}
 q^2 \quad \quad \quad \quad \quad for\quad d + 2 - \Gamma  - z > 0 \\
 q^{d + 2 - \Gamma  - z} \quad \quad \quad for\quad d + 2 - \Gamma  - z < 0 \\
 \end{array} \right.
\label{39},
\end{equation}
\begin{equation}
I_2^ <  \left( q \right) \propto \left\{ \begin{array}{l}
 const\quad \quad \quad \quad \quad for\quad d + 4 - 2\Gamma  - 2z\left( {1 - 2\phi } \right) > 0 \\
 q^{d + 4 - 2\Gamma  - z} \quad \quad \quad \quad for\quad d + 4 - 2\Gamma  - 2z\left( {1 - 2\phi } \right) < 0 \\
 \end{array} \right.
\label{40}.
\end{equation}

We consider now the upper-right quadrant of the $\left( {\Gamma
,z} \right)$ plane, where a solution may be expected. The lines $d
+ 2 - \Gamma  - z = 0$ and $d + 4 - 2\Gamma  - 2z\left( {1 -
2\phi } \right) = 0$ divide the quadrant into four sectors. We
investigate next each sector separately to decide whether a
solution of the equations (\ref{37})-(\ref{38}) can exist there
or not (in the limit of small $q$'s).

Sector $\alpha $ is defined by $d + 2 - \Gamma  - z > 0$ and $d +
4 - 2\Gamma  - 2z\left( {1 - 2\phi } \right) > 0$. In this sector
equations (\ref{37}) and (\ref{38}) reduce to
\begin{equation}
Aq^{ - \Gamma } \left[ {\left( {2\nu  + D_1  + D_2 } \right)q^2 -
Bq^z } \right] = \frac{{D_0 B^{ - 2\phi } }}{{A^\infty  \cos
\left( {\pi \phi } \right)\Gamma \left( {2\phi } \right)}}q^{ -
2\phi z}
\label{41},
\end{equation}
and
\begin{equation}
A\left( \nu  + E_1 +E_4\right)q^{2 - \Gamma } - E_5 - E_2  = \frac{{D_0 B^{ - 2\phi } }}{{\cos \left( {\pi \phi
} \right)}}q^{ - 2\phi z}
\label{42}.
\end{equation}

First, the possibility that $2-\Gamma>-2z\phi$ can be ruled out
immediately, because $B$ is positive, so that eq. (\ref{41})
cannot be balanced in leading order (in powers of $q$). If $2 -
\Gamma  <  - 2z\phi $ then the right-hand side of eq. (\ref{41})
is negligible compared to the left-hand side, so that the leading
order equations are identical to those obtained for the
white-noise KPZ problem and thus the standard KPZ results from
refs. \cite{SE98,katzav99} are restored. Therefore, we get $\Gamma
= 2$ and $z = 2$. Since $\Gamma = 2$ and $z$ must be positive, the
condition $2-\Gamma < -2z\phi$ can be met only for $\phi \leq 0$.
Namely, for the case of noise anticorrelated in time
\cite{schwartz93}. Such a solution holds only for $d>4(1-2\phi)$.

The other relevant option is $2 - \Gamma  =  - 2z\phi $. This
implies that $\phi$ must be positive, $\Gamma>2$ and $z \geq 2$.
There is now an interesting difference between the case $z>2$ and
$z=2$. For $z>2$, the leading order terms in eqs. (\ref{41}) and
(\ref{42}) lead to two linear homogeneous equations in the
quantities $A$ and $B^{-2\phi}$. This implies that in order to
have a physical solution with $A,B>0$, we must have the
determinant of the coefficient matrix vanish, namely
\begin{equation}
\Gamma \left( {2\phi } \right)A^\infty \left( {2\nu  + D_1  + D_2
} \right)=\left( \nu  + E_1 +E_4\right)
\label{42.5},
\end{equation}

Since the quantities $D_1,D_2,E_1,E_2$ depend on the behavior of
$\phi_\ell$ and $\omega_\ell$ for $\ell > q_0$, on the total upper
cutoff etc., it is difficult to envisage that eq. (\ref{42.5}) can
be fulfilled under accidentally, for non-generic values of the
parameters of the system. The case with $z=2$ is different. The
two equations for the coefficients $A$ and $B$ have now an
additional term - $AB$ on the right hand side of the first
equation. This enables now a generic solution for the
coefficients. In that case $\Gamma=2+4\phi$. Considering the
defining conditions for the sector we find that within the sector
such a solution is possible only for $d>4$.

Sector $\beta$ is defined by $d + 2 - \Gamma  - z > 0$ and $d + 4 - 2\Gamma  - 2z\left( {1 - 2\phi }\right)<0$.
In this sector equation (\ref{42}) is replaced by
\begin{equation}
A\left(\nu  + E_1 +E_4\right)q^{2 - \Gamma }- E_6q^{d+4-2\Gamma-z}
- E_2  = \frac{{D_0 B^{ - 2\phi } }}{{\cos \left( {\pi \phi }
\right)}}q^{ - 2\phi z}
\label{43},
\end{equation}
while equation (\ref{41}) remains intact.

The analysis for the possibility $2-\Gamma = -2z\phi$ in sector
$\beta$ is similar to the above analysis for sector $\alpha$. The
only difference is that due to the different defining conditions
of the sector, such a solution with $z=2$ and $\Gamma=2+4\phi$
holds within the sector for $2 + 4\phi < d < 4$.

Combining the results for sectors $\alpha $ and $\beta $, we see
that for $\phi  \le 0$ the noise term is irrelevant, and the
critical exponents that describe the EW problem with uncorrelated
noise (namely $\Gamma  = z = 2$) are restored. This option is
possible if $d>2$. In addition, for $\phi  > 0$ we get the new
solution $z = 2$ and $\Gamma  = 2 + 4\phi $ that is just the
solution obtained for the EW equation with temporally correlated
noise (see eq. (\ref{12}) above). Following the above discussion
it is realized that this solution is possible only for $d > 2 +
4\phi $. Therefore, the lower critical dimension in this problem
is $d_c = 2 + 4\phi $ (provided $\phi
> 0$, otherwise $d_c  = 2$ as mentioned above).

Sector $\gamma $ is defined by $d + 2 - \Gamma  - z < 0$ and $d +
4 - 2\Gamma  - 2z\left( {1 - 2\phi } \right) > 0$. In this sector
equation (\ref{42}) is replaced by
\begin{equation}
A\left( \nu  + E_1\right)q^{2 - \Gamma } + AE_7q^{d+4-2\Gamma-z}- E_5 - E_2  = \frac{{D_0 B^{ - 2\phi } }}{{\cos
\left( {\pi \phi } \right)}}q^{ - 2\phi z}
\label{44}.
\end{equation}

First, The two defining conditions of this sector imply that the first term on the left hand side is negligible
compared to the second term, and the second term is negligible compared to the term on the right hand side of
the equation. Therefore, looking at the simplified equation, we must conclude that $\phi = 0$ and $ - E_5 - E_2
= \frac{{D_0 B^{ - 2\phi } }}{{\cos \left( {\pi \phi } \right)}}$. However, this is impossible because the
left-hand side is negative definite.

Sector $\delta$ is defined by  $d + 2 - \Gamma  - z < 0$ and $d + 4 - 2\Gamma  - 2z\left( {1 - 2\phi } \right) <
0$. In this sector eqs. (\ref{37}) and (\ref{38}) take the form
\begin{equation}
Aq^{ - \Gamma } \left[ {\left( {2\nu  + D_1 } \right)q^2  - Bq^z + \frac{{8\lambda ^2 }}{{\left( {2\pi }
\right)^d }}\frac{A}{B}q^{d + 4 - \Gamma  - z}G\left( {\{f\},\Gamma ,z} \right) } \right] = \frac{{D_0 B^{ -
2\phi } }}{{A^\infty  \cos \left( {\pi \phi } \right)\Gamma \left( {2\phi } \right)}}q^{ - 2\phi z} \label{45},
\end{equation}
and
\begin{equation}
A\left(\nu  + E_1\right)q^{2 - \Gamma } - \frac{{2\lambda ^2 }}{{\left( {2\pi } \right)^d }}\frac{{A^2 }}{B}q^{d
+ 4 - 2\Gamma  - z}F\left( {\{f\},\Gamma ,z} \right)  - E_2  = \frac{{D_0 B^{ - 2\phi } }}{{\cos \left( {\pi
\phi } \right)}}q^{ - 2\phi z} \label{46}.
\end{equation}
where $G\left( {\{f\},\Gamma ,z} \right)$ is given by
\begin{equation}
G\left( {\{f\},\Gamma ,z} \right) = \int {d^d t\frac{{\left[ {\vec t \cdot \left( {\hat e - \vec t} \right)}
\right]\left[ {\hat e \cdot \left( {\hat e - \vec t} \right)} \right]}}{{\left| {\hat e - \vec t} \right|^z
}}\left| {\hat e - \vec t} \right|^{ - \Gamma } F_1 \left( {\left\{ f \right\},{\textstyle{{t^z } \over {\left|
{\hat e - \vec t} \right|^z }}}} \right)} \label{47},
\end{equation}
and $F\left( {\{f\},\Gamma ,z } \right)$ given by
\begin{eqnarray}
 F\left( {\{f\},\Gamma ,z} \right) =  &-& 4\int {d^d t\frac{{\left[ {\vec t \cdot \left( {\hat e - \vec t} \right)} \right]\left[ {\hat e \cdot \left( {\hat e - \vec t} \right)} \right]}}{{t^z }}\left| {\hat e - \vec t} \right|^{ - \Gamma } F_2 \left( {\left\{ f \right\},{\textstyle{{\left| {\hat e - \vec t} \right|^z } \over {t^z }}},{\textstyle{1 \over {t^z }}}} \right)}  \nonumber\\
&+& \int {d^d t\left[ {\vec t \cdot \left( {\hat e - \vec t} \right)} \right]^2 \left| {\hat e - \vec t}
\right|^{ - \Gamma } t^{ - \Gamma } F_3 \left( {\left\{ f \right\},\left| {\hat e - \vec t} \right|^z ,t^z }
\right)} \label{48}.
\end{eqnarray}
$\hat e$ is a unit vector in an arbitrary direction, and the
$\vec t$-integration is over all $d$-dimensional space.

From the defining conditions of this sector, it is possible to neglect the $q^2 $-term in the brackets on the
left-hand side of eq. (\ref{45}) compared to the third $q^{d + 4 - \Gamma  - z} $-term (since $d + 4 - \Gamma  -
z < 2$). In addition, It is also possible to neglect the $q^{2 - \Gamma }$-term on the left-hand side of eq.
(\ref{46}) compared to the second $q^{d + 4 - 2\Gamma  - z} $-term (since $d + 4 - 2\Gamma  - z < 2 - \Gamma $).
It is easy to see that for $\phi<0$ all the usual KPZ results are trivially retained, as the right hand side is
irrelevant then. Therefore, we focus on the case $\phi>0$ and thus the constant term on eq. (\ref{46}) can be
neglected compared to the noise term on the right hand side.

Now there are two options: First, when $\Gamma  > z\left( {1 + 2\phi } \right)$ then the right-hand side of both
equations (\ref{45})-(\ref{46}) is negligible compared to the left-hand side. In that case the critical
exponents are determined by a combination of the scaling relation $d + 4 - \Gamma  - 2z = 0$, and the equation
\begin{equation}
F\left[ {\{f\},\Gamma ,z\left( \Gamma  \right)} \right] = 0
\label{trans},
\end{equation}
(where $F$ is given by eq. (\ref{48}) above). We denote the solutions of the transcendental equation by $\Gamma
_\phi \left( d \right)$ (since the exponent $\Gamma$ is dependent on the spatial dimension $d$ and on $\phi$).
For example, in one dimension, and for $\phi=0$ it can be shown analytically that $\Gamma _0 \left( 1 \right) =
2$ and in two dimensions a numerical solution of the equation (again for $\phi=0$) yields $\Gamma _0 \left( 2
\right) = 2.59$ (see refs. \cite{SE98, katzav99}). A discussion for general $\phi$'s will be given below. Still,
we must remember that a solution here is obtained by requiring $\Gamma  > z\left( {1 + 2\phi } \right)$. This
yields a necessary condition for the existence of such a solution, $\phi  <  \frac{{3\Gamma _\phi \left( d
\right) - d - 4}}{{2\left( {d + 4 - \Gamma _\phi \left( d \right)} \right)}}$.

The second option in this sector is $\Gamma  = z\left( {1 + 2\phi } \right)$ (the possibility $\Gamma  < z\left(
{1 + 2\phi } \right)$ is irrelevant because one cannot balance the equations and still be consistent with the
defining conditions of sector $\delta$ in that situation). Then in order to balance eqs. (\ref{45})-(\ref{46})
we must also have $d + 4 - 2\Gamma - z =  - 2z\phi $. This leads to the new solution $z = \frac{{d + 4}}{{3 +
2\phi }}$ and $\Gamma = \left( {d + 4} \right)\frac{{1 + 2\phi }}{{3 + 2\phi }}$. However, this solution is
valid only if the equation (\ref{trans}) does not yield exponents (namely $\Gamma _\phi$ and $z_\phi $) that
make the $q^{d + 4 - 2\Gamma  - z}$-term in eq. (\ref{46}) dominant. Not surprisingly, this requirement
translates into the condition $\phi  >  \frac{{3\Gamma _\phi \left( d \right) - d - 4}}{{2\left( {d + 4 - \Gamma
_\phi \left( d \right)} \right)}}$ - implying either a smooth transition between the two types of solutions, or
 a complete domination of the first option (i.e. $\Gamma_\phi(d)$). Actually, the existence of this new solution
 also requires $F\left[ {\{f\},\Gamma ,z\left( \Gamma  \right)} \right] < 0$. This requirement turns out to be the same as
$\Gamma_\phi(d)<\Gamma_{new}=\left( {d + 4} \right)\frac{{1 + 2\phi }}{{3 + 2\phi }}$, so that this extra
requirement is fulfilled automatically since $\Gamma  = z\left( {1 + 2\phi } \right)$.

To summarize the results of sector $\delta$, we found two possible strong-coupling solutions. The first solution
is obtained from the equation (\ref{trans}), and its scaling exponents are denoted by $\Gamma_\phi(d)$ and
$z_\phi(d)$ (this solution reduces to the standard KPZ results when $\phi=0$). The second solution, is given by
the explicit expressions $\Gamma = \left( {d + 4} \right)\frac{{1 + 2\phi }}{{3 + 2\phi }}$ and $z = \frac{{d +
4}}{{3 + 2\phi }}$. Then, in a given dimension $d$ and for a given $\phi$, the actual strong coupling exponents
of the KPZ problem with temporally correlated noise are just $\Gamma=max \{\Gamma_\phi(d), \left( {d + 4}
\right)\frac{{1 + 2\phi }}{{3 + 2\phi }}\}$ and its corresponding $z$. Thus, the transition between the two
solutions as a function of $\phi$ (if such a transition exists) is continuous. However, it should be emphasized
that for a specific $\phi$ one of these solutions dominates so there is no phase transition between them.

Based on the results of the linear theory (\ref{12}) and the
numerical simulation \cite{Lam92} we expect the exponent $\Gamma$
to be a nondecreasing function of $\phi$ (that means that the
inclusion of temporal correlations does not make the surface
smoother). This implies that $\Gamma_\phi (d) \ge \Gamma_0 (d)$,
and since we know $\Gamma_0 (d)$ from the white-noise KPZ problem
it is easy to determine a lower bound on $\phi$, denoted by
$\phi_c(d)$, such that the new solution $\Gamma = \left( {d + 4}
\right)\frac{{1 + 2\phi }}{{3 + 2\phi }}$ is not possible below
it. The specific value of $\phi_c(d)$ is $\phi_c(d)=\frac{{3\Gamma
_0 \left( d \right) - d - 4}}{{2\left( {d + 4 - \Gamma _0 \left( d
\right)} \right)}}$. The fact that $\Gamma_\phi(d)$ is
non-decreasing as a function of $\phi$ implies two possible
options. Either the expression $\Gamma_\phi(d)$ gives the strong
coupling solution for the whole range of $0<\phi<1/2$, or it is
the solution for small $\phi$ and crosses over to $\Gamma = \left(
{d + 4} \right)\frac{{1 + 2\phi }}{{3 + 2\phi }}$. Such a
crossover can occur only above $\phi_c(d)$.

We turn now to the actual evaluation of the exponents. To do that
we need an ansatz for the scaling function  $f$ on which the form
of the equations for $\Gamma$ and $z$ depend. Since the equations
were constructed in such a way that second order corrections to
the quantities $\phi_q$ and $\omega_q$ vanish, we use as in refs.
\cite{SE02,SE02b}, the zero order form of the scaling function in
evaluating these corrections. The scaling function in zero order
is the function obtained for the corresponding linear theory,
given by eq. (\ref{12.4}). We therefore simplify eq. (\ref{trans})
using this ansatz. First, one quantity can be evaluated exactly
\begin{equation}
F_1 \left( {\left\{ {f_{EW} } \right\},a} \right) = \int_0^\infty  {e^{ - ax} f_{EW} \left( x \right)dx} =
\frac{{a - a^{ - 2\phi } }}{{a^2  - 1}} \label{48.2}.
\end{equation}
In addition, the functional $F_2$ can be simplified so that it involves only one dimensional integration
(instead of double integration)
\begin{eqnarray}
F_2 \left( {\left\{ {f_{EW} } \right\},a,b} \right) &=& \int\limits_0^\infty  {dx\int\limits_0^\infty  {dye^{ - x - y} f_{EW} \left( ax \right)f_{EW} \left( bx + y \right)} }  \nonumber\\
&=& \frac{{\cos \left( {\pi \phi } \right)}}{\pi }\int\limits_{ - \infty }^\infty  {du\frac{{\left| u \right|^{
- 2\phi } }}{{u^2  + 1}}\frac{1}{{1 - iu}}} \frac{{a^{1 + 2\phi } \left( {1 - ibu} \right)^{ - 2\phi } - \left(
{1 - ibu} \right)}}{{\left[ {a^2  - \left( {1 - ibu} \right)^2 } \right]}} \label{48.3},
\end{eqnarray}
(note that the integral is real, as it should be, even though the integrand is complex).

In order to illustrate the outcome of this analysis we specialize
to one dimension. First, in one-dimension only strong coupling
solutions are possible as the critical dimension is $2+4\phi$.
Second, as mentioned above, in one dimension eq. (\ref{trans}) can
be solved analytically for $\phi=0$ and it gives $\Gamma _0 = 2$.
This corresponds to a roughness exponent of $\alpha _0 = {1
\mathord{\left/ {\vphantom {1 2}} \right.
\kern-\nulldelimiterspace} 2}$ (using eq. (\ref{36})) and to a
dynamic exponent of $z_0  = {3 \mathord{\left/ {\vphantom {3 2}}
\right. \kern-\nulldelimiterspace} 2}$ (using the scaling relation
$z\left( \Gamma  \right) = {{\left( {d + 4 - \Gamma } \right)}
\mathord{\left/ {\vphantom {{\left( {d + 4 - \Gamma } \right)} 2}}
\right. \kern-\nulldelimiterspace} 2}$). for higher values of
$\phi$ one has to solve equation (\ref{trans}) numerically using
the ansatz of the linear theory. These results are summarized in
Fig. \ref{exps} as the solid line. The figure also presents the
possible second solution $\Gamma  = 5{{\left( {1 + 2\phi }
\right)} \mathord{\left/ {\vphantom {{\left( {1 + 2\phi } \right)}
{\left( {3 + 2\phi } \right)}}} \right. \kern-\nulldelimiterspace}
{\left( {3 + 2\phi } \right)}}$ (that corresponds to $\alpha  =
{{\left( {1 + 4\phi } \right)} \mathord{\left/ {\vphantom {{\left(
{1 + 4\phi } \right)} {\left( {3 + 2\phi } \right)}}} \right.
\kern-\nulldelimiterspace} {\left( {3 + 2\phi } \right)}}$) and $z
= {5 \mathord{\left/ {\vphantom {5 {\left( {3 + 2\phi } \right)}}}
\right. \kern-\nulldelimiterspace} {\left( {3 + 2\phi }
\right)}}$, and a continuation of $\Gamma_0$ as a dashed line.
However, since the this solution is smaller than $\Gamma_\phi$, it
is practically irrelevant, since $\Gamma_\phi$ dominates the whole
$\phi$ range.

\begin{figure}[htb]
\includegraphics[width=8cm]{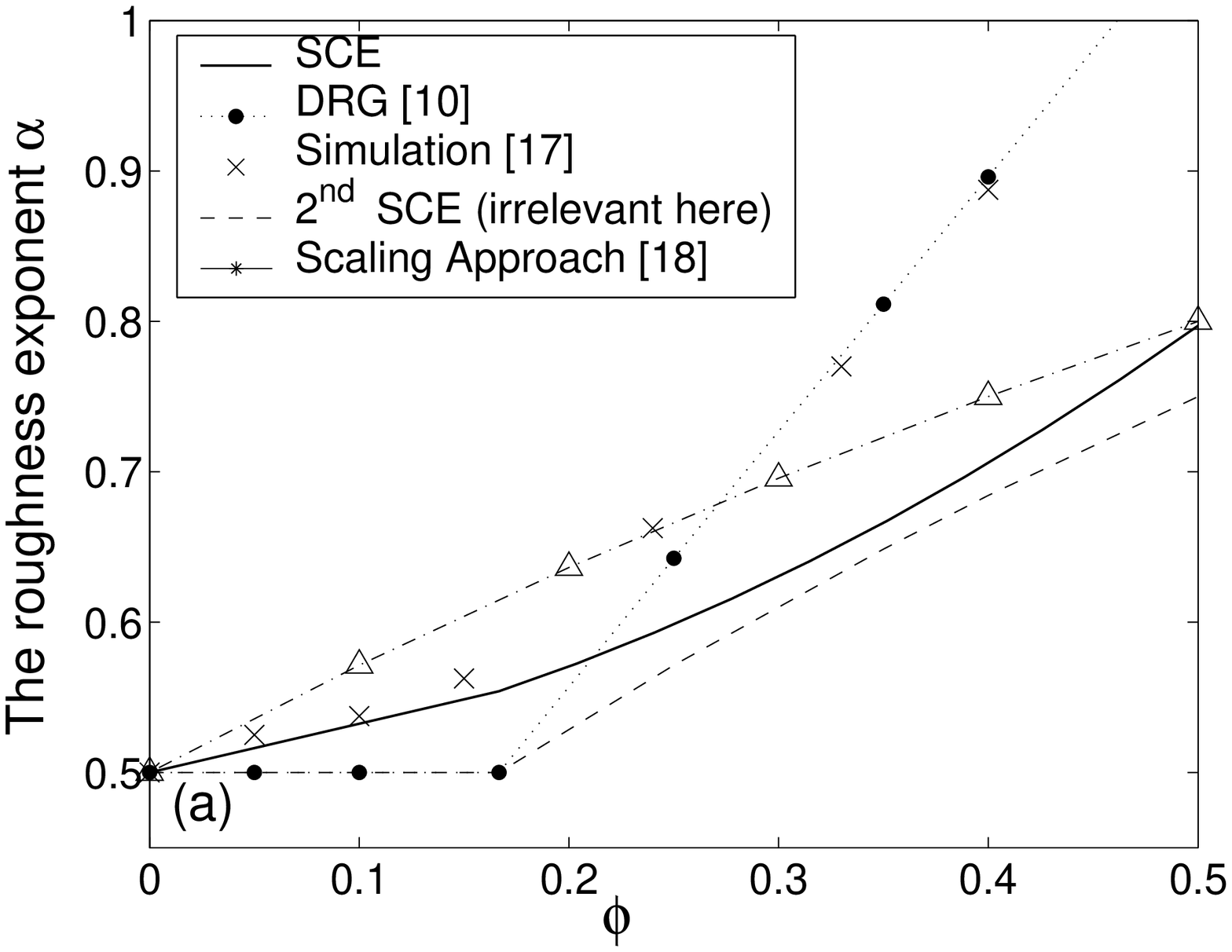}
\includegraphics[width=8cm]{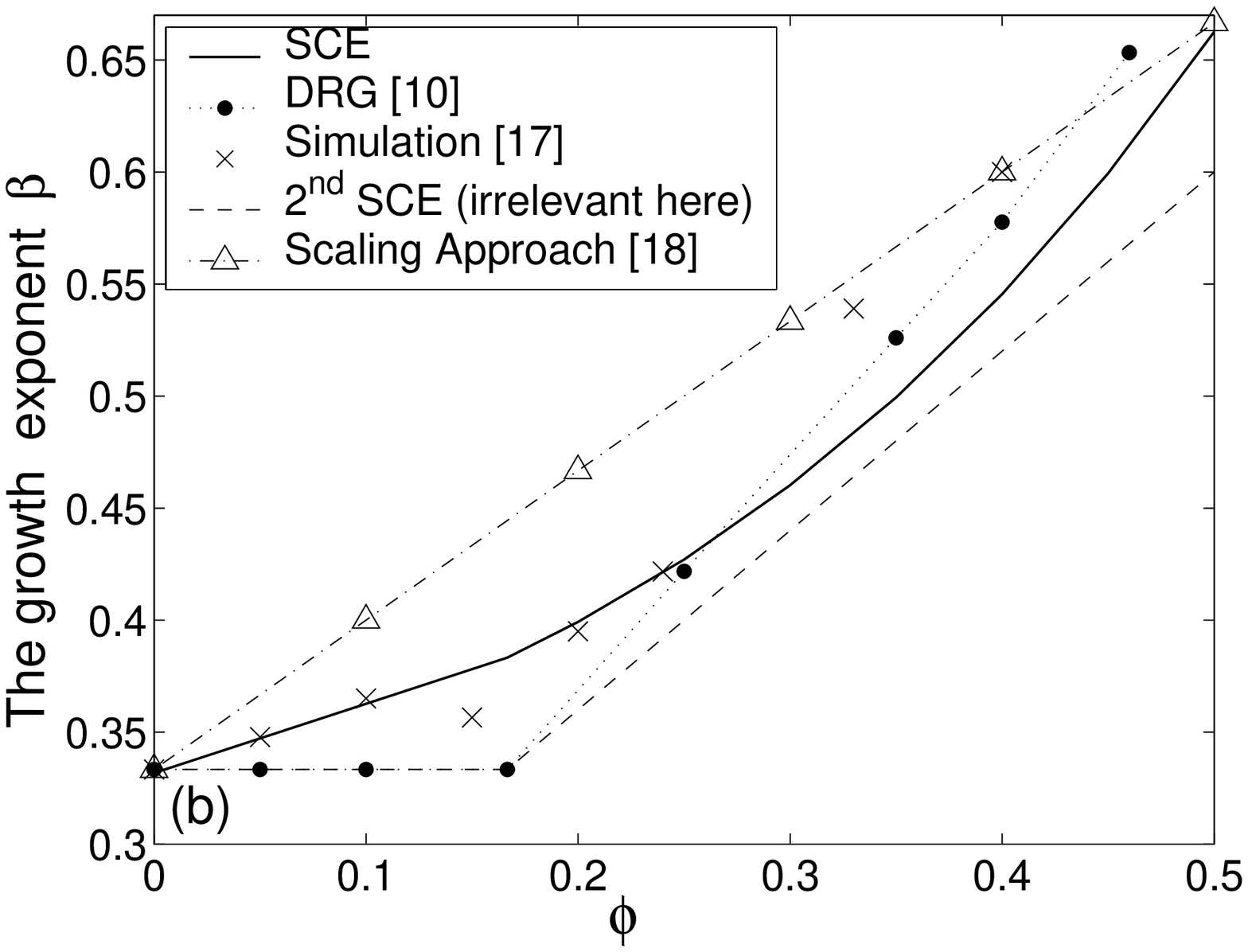}
\includegraphics[width=8cm]{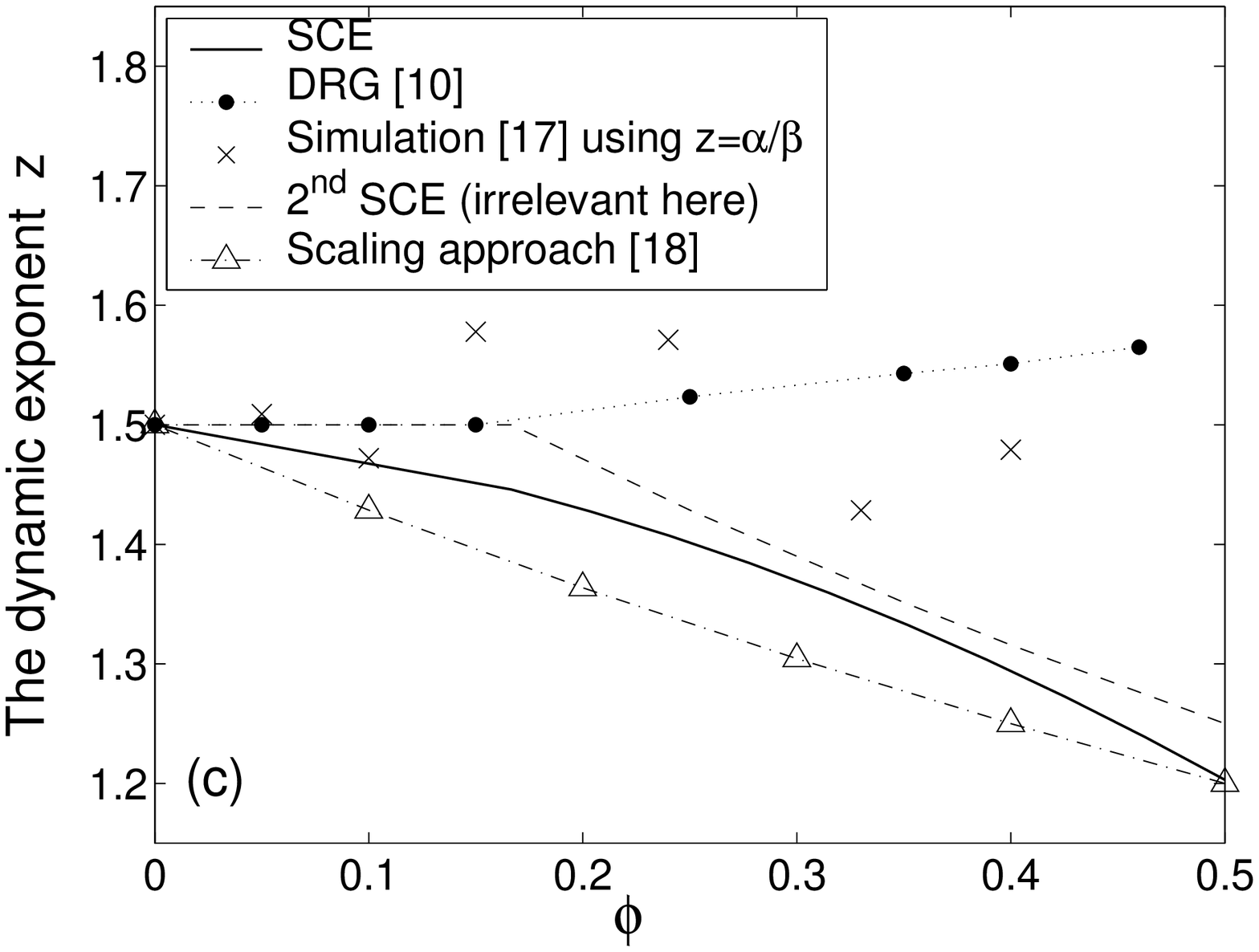}
\caption{(a) The roughness exponent $\alpha_\phi$, (b) the growth exponent $\beta_\phi$ and (c) the dynamic
exponent $z_\phi$ as a function of the exponent $\phi$ for decay of temporal correlations in $d =1$. Note that
the dynamic exponent $z$ was inferred for the numerical results of ref. \cite{Lam92} from $\alpha$ and $\beta$
using the scaling relation $z=\alpha/\beta$. Second, notice that the DRG result is possible only up to
$\phi=0.46$. Third, the dashed line shows our second possible solution (using SCE) that turns out to be
irrelevant here, since it is smaller than the SCE $\alpha_\phi$ for all $\phi$'s.} \label{exps}
\end{figure}

It is particularly interesting to compare this one-dimensional result to the DRG result that was presented in
the introduction \cite{medina89}, and plotted for convenience in Fig. \ref{exps}. Generally speaking, the two
methods disagree on the values of the critical exponents significantly over most of the $\phi$-range. Three
substantial differences can be observed between the two methods. First, using the self-consistent approach we
found no "threshold behavior". That is, we found a continuous variation of the scaling exponents $\alpha$ and
$z$ as a function of $\phi$ over the whole range of possible $\phi$'s, rather than no variation of these
exponents up to a critical value of $\phi_c$ and a quasi-linear behavior from that point on.

Second, we found a solution for the exponents for every $\phi$,
while the DRG approach found no solution above $\phi=0.46$
(claiming that no stable surface can grow under the condition
$0.46<\phi<0.5$). Interestingly, the threshold $\phi_c$ (the
crossover point) that was predicted using DRG in $d=1$ (namely
$0.167$) is the same as the lower bound we found above
($\phi_c=1/6$ for $d=1$). Our exact statement was that for
$\phi>1/6$ the second strong-coupling KPZ solution becomes
possible in principle (but not in practice). Therefore, the DRG
result might reflect this exact statement.

Third, we found that $z$ is a decreasing function of $\phi$, while
the DRG approach predicts an increasing value of $z$. The reason
for this difference is not clear, but it might stem from the
definition of the typical "decay rate", that was actually defined
using the scaling form (\ref{17.5}), rather than a more
"traditional" definition such as eq. (\ref{13}). The reason for
using this definition is that the integral over the scaling
function $f(u)$ does not converge, because of its power-law tail.
Actually, in the case of the linear theory, where everything can
be calculated exactly, the only possible definition is the one we
used. Now, since the introduction of temporally correlated noise
certainly slows down the relaxations in the system, this might
have caused an artifact of increasing $z$, because larger $z$'s
are interpreted as longer relaxation times. However, in our
approach, we do see this slowing down clearly, but it does not
come from a larger dynamic exponent $z$ in an exponential decaying
scaling function, but rather from a very slowly decaying scaling
function, which does not decay exponentially. Thus, this
difference might reflect a better understanding of the
time-dependent dynamics in such driven systems.

The discrepancy between DRG and our result is obvious. This should
not come as a surprise as it is not new that SCE and DRG yield
different results, apart from some special cases, like the
one-dimensional white noise (in space and in time) case. In fact
it should not be worrying in the present case because in many
cases studied in the past the results of SCE are either dimilar of
those of DRG (e.g., the one dimensional case with no correlations
\cite{kpz86, SE98}), or superior to them (as in the case of white
noise KPZ in higher dimensions \cite{SE98}, or in the case of the
nonlocal KPZ equation that is exactly soluble in one dimension
\cite{katzav02, katzav03}). On that basis we may expect this to be
true also here. This expectation is supported, in fact, by the non
physical feature of increasing $z$ predicted by DRG (see Fig.
\ref{exps}).

We must admit that we were surprised by the fact that that our
results do not show a threshold behavior (like DRG). Namely, the
fact that we do not find the characteristic exponents for a range
of small $\phi$'s to be identical to those found at $\phi=0$. In
fact, it is interesting that such a solution to the SCE equations
exists but it is only potentially possible. the actual solution
that determines the exponents is the second solution (in one
dimension). The characteristic exponents obtained by the scaling
approach \cite{Ma93} and by the numerical simulations do not
exhibit a threshold behavior as well (see Fig. \ref{exps}). A
closer inspections of the values of the results of the scaling
approach, reveals that these results are actually a simple
interpolation (linear for $\beta$ and almost linear for $\alpha$
and $z$) of the values of the exponents between the results we
predict for $\phi=0$ and $\phi=1/2$.

The results of the simulations are even more interesting. The
simulations \cite{Lam92} reports the values of $\alpha$ and
$\beta$ as a function of $\phi$ (these values are recovered in
Fig. \ref{exps}). The first four small-$\phi$ points for $\alpha$
and the first three small-$\phi$ points for $\beta$ agree with our
curves. Thus, the simulations and SCE, which are completely
independent, agree exactly in the region of $\phi$ where threshold
behavior could have been expected. This suggest that the small
$\phi$ behavior predicted by both methods is correct and indeed a
threshold behavior should not be expected here. For higher
$\phi$'s the results of simulations obviously deviate from our
curves. Can the results of simulations be trusted for large
$\phi$'s? In their paper Lam and Sander \cite{Lam92} report
difficulties observed for larger $\phi$'s, but they also report
measures taken to ensure the correctness of their final result. To
check whether indeed those larger $\phi$ results could be trusted,
we took the freedom of drawing the dynamical exponent
$z=\alpha/\beta$ as inferred from the simulations that give only
$\alpha$ and $\beta$ (the inferred values of $z$ are presented in
Fig. \ref{exps}(c) above). The erratic oscillations of the dynamic
exponent obtained from the simulation, strongly suggest that the
measures taken by the authors to eliminate the observed larger
$\phi$ problems were probably not enough.

\section{GENERALIZATION TO SPATIO-TEMPORALLY CORRELATED NOISE}

Up to this point we discussed the relatively simple case of noise
without any spatial correlations (i.e. $D^0 \left( q \right) =
D_0 $). However, including spatial correlations bears no
principal difficulty to the analysis presented above. For example
one can just replace $D_0 $ with $D^0 \left( q \right) = D_0 q^{
- 2\rho } $ from eq. (\ref{41}) and on, and thus can easily
obtain the scaling exponents for that case. For the sake of
presenting a complete picture we briefly summarize the results
obtained for this case. First, there is the weak-coupling
solution, which is again just the corresponding EW result for
such a noise, given by
\begin{equation}
z = 2\quad \quad and\quad \quad \Gamma  = 2 + 2\rho  + 4\phi
\label{49}.
\end{equation}
The weak-coupling solution is possible for $d > 2 + 2\rho  +
4\phi $, so that here the lower critical dimension is $d_c  = 2 +
2\rho  + 4\phi $.

Second, there is the strong coupling solution, given by
\begin{equation}
z = \left\{ \begin{array}{l}
 z_\phi \left( d \right) \\
 \frac{{d + 4 - 2\rho }}{{3 + 2\phi }} \\
 \end{array} \right.\quad and\quad \Gamma  = \left\{ \begin{array}{l}
 \Gamma _\phi \left( d \right) \quad \quad \quad \quad \quad 2\rho  + \left( {1 + 2\phi } \right)z_\phi \left( d \right) < \Gamma _\phi \left( d \right) \\
 \frac{{(d + 4 - 2\rho)(1 + 2\phi)}}{{3 + 2\phi }}\quad \quad 2\rho  + \left( {1 + 2\phi } \right)z_\phi \left( d \right) > \Gamma _\phi \left( d \right) \\
 \end{array} \right.
\label{50},
\end{equation}
where as before, $\Gamma _\phi \left( d \right)$ is the solution of the equation (\ref{trans}) and $z_\phi
\left( d \right) = {{\left( {d + 4 - \Gamma _\phi \left( d \right)} \right)} \mathord{\left/ {\vphantom {{\left(
{d + 4 - \Gamma _0 \left( d \right)} \right)} 2}} \right. \kern-\nulldelimiterspace} 2}$.

Furthermore, the method presented above is not restricted to noise terms that have separable correlators, i.e.
$D\left( {q,\omega } \right) = D^0 \left( q \right)\omega ^{ - 2\phi }$, and can just as well deal with
non-separable correlators (that is any functional form of $D\left( {q,\omega } \right)$). In that case, the only
difference would be to replace the right-hand side of eq. (\ref{20}) by the expression $\frac{{D\left( {q,\omega
} \right) }}{{\omega ^2 + \omega _q^2 }} $ with the required $D\left( {q,\omega } \right)$ inside.

In order to demonstrate this option, we discuss an interesting
application of this approach to the KPZ equation with a very
special kind of spatio-temporally correlated noise (this result
was mentioned at the end of the introduction). This problem was
previously solved by Li et al. \cite{Li96} in the context of
Vortex lines in the three-dimensional XY model with random phase
shifts, and it boils down to solving the KPZ equation in two
dimensions (more specifically the $2 + 1$ case) with a noise term
that has the following spatio-temporal correlations
\begin{equation}
\left\langle {\eta \left( {\vec r,t} \right)\eta \left( {\vec
r',t'} \right)} \right\rangle  = \frac{{\sigma \left( {Jm}
\right)^2 }}{{\sqrt {\left( {\vec r - \vec r'} \right)^2  +
\left( {t - t'} \right)^2 } }}
\label{51},
\end{equation}
where $\vec r$ is a two dimensional vector (i.e. $\vec r = \left(
{x,y} \right) \in R^2 $). In order to apply the method presented
above for finding the critical exponents of this model we have to
Fourier transform the noise correlator first
\begin{equation}
\left\langle {\eta \left( {\vec q,\omega } \right)\eta \left(
{\vec q',\omega '} \right)} \right\rangle  = \sigma \left( {Jm}
\right)^2 \frac{{\delta ^2 \left( {\vec q + \vec q'}
\right)\delta \left( {\omega  + \omega '} \right)}}{{q^2  +
\omega ^2 }}
\label{52},
\end{equation}
(where $\vec q$ is a two-dimensional vector) so that $D\left(
{q,\omega } \right) = \frac{{D_0 }}{{q^2  + \omega ^2 }}$ (with
$D_0  \equiv \sigma \left( {Jm} \right)^2$). Now, for small $q$'s
$\omega _q  = Bq^z $, and we can evaluate the Fourier integral
explicitly (assuming $z>1$)
\begin{equation}
\int\limits_{ - \infty }^\infty  {\frac{{e^{i\omega t} }}{{\left(
{\omega ^2  + \omega _q^2 } \right)\left( {\omega ^2  + q^2 }
\right)}}d\omega } \sim \frac{1}{{\omega _q^3 }}\frac{\pi
}{2}\frac{{e^{ - \omega _q t} }}{{{{q^2 } \mathord{\left/
 {\vphantom {{q^2 } {\omega _q^2 }}} \right.
 \kern-\nulldelimiterspace} {\omega _q^2 }}}} = \frac{1}{{q^{2 + z} }}\frac{{\pi e^{ - \omega _q t} }}{{2B}}
\label{53}.
\end{equation}
As mentioned above, this expression should replace the right-hand side of eqs. (\ref{22}) and (\ref{27}). One
can easily be convinced that this term is subdominant in the long-time limit, and therefore drops out of eq.
(\ref{25}). However, in the short-time limit this term is not negligible, and therefore modifies eq. (\ref{28})
accordingly.

Following all the required steps from that point on leads to the results $\Gamma  = z + 2$ and $d + 4 - 2z -
\Gamma  = 0$, so that for the case we were interested in $\left( {d = 2} \right)$ we get
\begin{equation}
z = \frac{4}{3}\quad \quad and\quad \quad \Gamma  = \frac{{10}}{3}
\label{54},
\end{equation}
identical to the analytic result presented in ref. \cite{Li96} (a
numerical analysis presented there also verifies this result).

\section{SUMMARY AND CONCLUSIONS}

In this paper we developed a time-dependent self-consistent
approach to deal with the KPZ equation driven by a temporally
correlated noise. This achievement was made possible thanks to
the observation that there is a time scale separation between
short-time and long-time behavior of the system. More
specifically, it was realized that when temporally correlated
noise is present in the system, then slow relaxations of various
time-dependent quantities should control the long time behavior
(in this case algebraic decay of the time-dependent correlation
function $\Phi _q \left( t \right)$). In addition, it was seen
that the short time behavior is influenced by the long time
behavior and vise versa.

To summarize the results briefly, we found that the KPZ equation
with temporally correlated  noise, just like the problem with
white noise, has both a strong-coupling and a weak-coupling
solution. The weak coupling solution is described by the scaling
exponents of the corresponding linear theory (EW equation), and is
made possible for dimensions higher than the lower critical
dimension $d_c=2+4\phi$ (the specific values are given in eq.
(\ref{49}). The strong-coupling solution, which is relevant also
for low dimensions, is described by critical exponents that are a
result of a competition between two possible solutions. First,
there is an extension of the classical white-noise KPZ solution
denoted by $\Gamma_\phi(d)$ that is derived from an integral
equation (\ref{trans}). The other possible solution is a new
strong-coupling solution and is given in eq. (\ref{50}). The
actual exponent $\Gamma$ that describes the surface is the maximum
between the two options. For small $\phi$'s $\Gamma_\phi(d)$ is
the solution but it may (or may not, as in  the one dimensional
case) cross over in a continuous manner to the second solution.
For a detailed discussion and comparison to other methods in one
dimension see section V.

The comparison suggest that for small values of $\phi$ our results
are correct, being supported by the totally independent numerical
simulations. For larger values of $\phi$, the erratic behavior of
$z$ obtained from the simulations suggests a problem in the
numerical evaluation of the scaling exponents. Therefore, an
independent study, preferably a heavy numerical study, aimed at
larger $\phi$'s could prove useful in a critical determination of
the scaling exponents.

\end{document}